# Epistemology of Modeling and Simulation:
# How can we gain Knowledge from Simulations?

| | |
|---|---|
| **Andreas Tolk** | **Saikou Y. Diallo; Jose Padilla; Ross Gore** |
| EMSE, Old Dominion University | VMASC, Old Dominion University |
| Norfolk, Virginia | Suffolk, Virginia |
| atolk@odu.edu | sdiallo@odu.edu; jpadilla@odu.edu; rgore@odu.edu |

## ABSTRACT

Epistemology is the branch of philosophy that deals with gaining knowledge. It is closely related to ontology. The branch that deals with questions like "What is real?" and "What do we know?" as it provides these components. When using modeling and simulation, we usually imply that we are doing so to either apply knowledge, in particular when we are using them for training and teaching, or that we want to gain new knowledge, for example when doing analysis or conducting virtual experiments. This paper looks at the history of science to give a context to better cope with the question, how we can gain knowledge from simulation. It addresses aspects of computability and the general underlying mathematics, and applies the findings to validation and verification and development of federations. As simulations are understood as computable executable hypotheses, validation can be understood as hypothesis testing and theory building. The mathematical framework allows furthermore addressing some challenges when developing federations and the potential introduction of contradictions when composing different theories, as they are represented by the federated simulation systems.

## ABOUT THE AUTHORS

**Andreas Tolk** is Professor of Engineering Management and Systems Engineering with a joint appointment to Modeling, Simulation, and Visualization Engineering at Old Dominion University. He is also affiliated with the Virginia Modeling Analysis and Simulation Center. He holds a M.S. and Ph.D. in Computer Science from the University of the Federal Armed Forces in Munich, Germany.

**Saikou Y. Diallo** is Research Assistant Professor at the Virginia Modeling, Analysis and Simulation Center at Old Dominion University. He received a B.S. in Computer Engineering, and a M.S and Ph.D. in Modeling and Simulation from Old Dominion University.

**Jose J. Padilla** is Research Assistant Professor at the Virginia Modeling, Analysis and Simulation Center at Old Dominion University. He received a B.S. in Industrial Engineering from the Universidad Nacional de Colombia, an M.B.A. in International Business from Lynn University, and a Ph.D. in Engineering Management from Old Dominion University.

**Ross Gore** is a Post-Doctoral Research Associate at Virginia, Modeling Analysis and Simulation Center at Old Dominion University. He holds a B.S. in Computer Science from the University of Richmond, and a M.S. and PhD. In Computer Science from the University of Virginia.

# Epistemology of Modeling and Simulation: How can we gain Knowledge from Simulations?

**Andreas Tolk**
EMSE, Old Dominion University
Norfolk, Virginia
atolk@odu.edu

**Saikou Y. Diallo; Jose Padilla; Ross Gore**
VMASC, Old Dominion University
Suffolk, Virginia
sdiallo@odu.edu; jpadilla@odu.edu; rgore@odu.edu

## INTRODUCTION

The quote that essentially *"All models are wrong, but some are useful!"* is attributed to George E.P. Box. This quote raises the question if we can gain knowledge from something that is essentially wrong? What are we doing when we build models, derive simulations, and then execute the simulations? What is the scientific justification for applying models? Is simulation really the new third column of science as suggested in the NSF Report (2006), standing as an equal partner beside theory and experimentation?

Within this paper, we will look at several foundations for Modeling and Simulation (M&S) to justify why we can learn from M&S despite a lot of challenges connected with the epistemology of M&S, which is dealing with the theory of the nature and grounds of knowledge gained with M&S especially with reference to its limits and validity.

We will start with a very limited view on science and how we gain understanding using the scientific method. As the category of M&S as we are in particular interested in is computer-based simulation, we will than look into computability next.

Finally, we will look at the mathematical foundations of M&S, in particular when it comes to the development of federations, i.e., more than one independently developed simulation systems are combined using interoperability protocols to support a common objective.

All these observations lead to the need to revisit our ideas of verification and validation to ensure that we not only learn "something" from M&S, but that we gain knowledge in the epistemological sense.

## A BRIEF HISTORY OF SCIENCE

It may seem to be unnecessary to many readers starting this paper with a brief history of science, as every young student learns the scientific principles in elementary school already: In order to understand a phenomenon the scientist formulates a hypothesis that predicts the outcome of an experiment. If the outcome is actually observed, the hypothesis is supported and becomes an explanation and ultimately can contribute to a theory.

This understanding, however, is actually not very old, and actually is not even one common understanding. Goldman (2006) provides some fundamental insights. It goes beyond the scope of this paper to deal with the details, but the following observations prepare the ground for the arguments used later:

- The British jurist and educational reformer Francis Bacon is generally recognized as the father of the experimental method. He proposed a strictly controlled inductive-empirical method to gain knowledge: only based on observation and data collection, analyses of data can uncover correlations that lead to hypotheses, further testing, confirmation, and ultimately the recognition of nature's law.

- The French mathematician and philosopher René Descartes proposed at the same time an alternative approach. He supported the deductive-rational method. He put the mathematical model in the center and used deductive reasoning to get new insights. Experiments are just a tool of limited value, as all experiments had to be conducted using too many constraints and their results were often equivocal.

- Isaac Newton's work on the *"Mathematical Principles of Natural Philosophy"* (1687) became the foundation for scientific work for two and a half centuries. His laws could not be deducted from experiments, but were consistent with experience. However, he defined the components of his laws in way they supported his mathematics to describe the laws. They were neither inductively nor deductively derivable from experience or experimentation, but they allowed a coherent and consistent interpretation: a conceptualization of experience!

Today, we know that many of Newton's concepts are "wrong" – or better stated, they are not generally applicable – as relativity theory and quantum physics did lead to new insights and provided alternative views. These new concepts are the best we currently have, but how can we assume that these new conceptualizations are correct and will stand the test of time (and if, for how long)? Nonetheless, the laws described by Newton were very successfully applied to gain a better understanding of nature and, although no longer believed to be universally applicable, they are still fundamental in today's school education. Popper (1935) formulated the implications of these observations that theories cannot be proven to be generally correct, but we can state that they have not been falsified so far by new observations or insights.

Science is still defined by analyses of data, discovering of correlations, and explaining the correlations by causalities that have to be confirmed by observation and experimentation. New tools allow more accurate observations. New conceptualizations provide better structures to capture the knowledge in coherent frameworks … but nobody knows for sure that our current framework is the ultimate truth, or if it is just the best thing currently available.

The essence of this little discourse into the history of science shall be *that science itself is a series of models that provide functional causalities leading to observable correlations.* Although wrong from the current perspective, they all help mankind to gain new understanding about nature or, in other words, to produce knowledge that could be applied to solve problems and provide solutions.

## COMPUTABILITY AND M&S

In this paper, modeling is understood as the process to develop a model while simulation is understood as executing such a model. It is therefore understood that modeling resides on the abstraction level whereas simulation resides on the implementation level. Modeling identifies the data and the correlation, and provides the functionality representing the causality leading to the observed correlation. This is done by simplifying and abstracting from the observation. As such, a model is a purposeful, task-driven simplification and abstraction of a perception of reality. Each simulation is therefore an implementation of a model, which hopefully is explicitly captured in form of a conceptual model (Balci and Ormsby, 2007).

The authors are in particular interested in computer based simulations, so the aspect of computability becomes important. Computability deals with the question if something can be executed on a digital computer, and in particular whether the functions can be implemented as a computer program.

A function is computable when a finite algorithm exists that describes what the function is doing, and if the function works on a discrete and finite set of arguments. Many alternative but equivalent definitions exist that all use different models of computation, but the idea of discrete and limited range and domain and the existence of an algorithm are common features.

Simulations can become very complex and in recent history are equipped with impressive means of visualization. We can create breath-taking virtual worlds that seem as real as reality, but all these cannot take away the fact that computer-based simulations are made up out of computable functions, which means all the constraints and limitations of computable functions are applicable to simulations as well.

One of these constraints is that certain classes of problems cannot have an algorithmic solution. Turing (1936) described the halting problem as an example and gave the proof that no algorithm can exist that solves the problem. The problem is the following: "Given a program and an input to the program, determine if the program will eventually stop when it is given that input." The well-known proof works with the assumption that such an algorithm exists and derives a contradiction. As the only assumption is that such an algorithm exists, the contradiction is the proof that such an algorithm cannot exist, as it would lead to the contradiction:

> *Let us assume we can write the algorithm that solves the halting problem. We call it the program $H$. When the program halts, $H$ produces 'halt,' otherwise 'loop.'*
> *We use $H$ to construct the program $K$. $K$ uses $H$ and halts, when $H$ produces the output 'loop,' and when $H$ produces 'halt,' it loops forever.*
> *We can now apply $H$ to $K$. If $H$ says that $K$ halts then $K$ itself would loop, and if $H$ says that $K$ loops then $K$ will halt. This cannot be; which means $H$ cannot exist.*

The observation that many problems exist that cannot be decided by an algorithm is directly applicable to M&S, as these problems cannot be solved by M&S either. Known problems belonging to this class are questions like "Will the system terminate?", "Are two modeled actions order independent or do I have to orchestrate them?", "Is the specification complete?", "Is the specification minimal?", or "Are two specifications functionally equivalent, in other words, do the delivery the same functionality?"

Beside these challenges of existence of an effective algorithm (decidability of a problem) the challenge of finding an efficient algorithm (complexity of a problem) needs to be addressed as well. Even if we can solve a problem effectively, the time needed to solve it may be prohibitive to use such a solution in a simulation system. Only effective and efficient algorithms make sense for computer-based simulations, which limits the number of useful functions significantly in comparison with all functions that can be used to explain correlations.

It is worth to point out that at no point in this argument we limited it to a certain paradigm. The observations are true for Monte-Carlo simulations, continuous simulations, discrete event simulation, agent based simulation, and all other imaginable forms of simulation people may come up with in the future. The epistemological constraints of composability are universal for computer-based simulation.

## MATHEMATICAL FOUNDATIONS

Another aspect becomes obvious when looking at the mathematical foundations. A simulation system is a production system that applies the procedures captured in the system to transform input parameters into output parameters. The simulation itself can therefore be understood as a formal language that produces terms (output parameters) based on observations (input parameters). This justifies the application of model theory (Weiss and D'Mello, 1997).

Model theory is a branch of mathematics that deals with the interpretation of formal languages using set-theoretic structures. In particular, it deals with the equivalency of interpretations in different formal languages. The fundamental terms of model theory are the 'formal' *languages* that are used to express the concepts to be evaluated. In order to interpret a *sentence* of a language, a *structure* is needed. This structure interprets sentences to be true or false. The set of sentences that are interpreted to be true builds the *theory*. Structures are therefore understood as the *model* of a language. The theories of these models are the sets of true statements in these models. The detailed definitions and applications in support of M&S are given in Tolk et al. (2013).

We know from the earlier section on computability that no general algorithm can exist that decides if two formal system specifications are functionally equivalent.

However, by using model theory to define M&S allows the applications of findings and results already generally proven to gain new insights. E.g., two simulation systems are equivalent if they produce the same sentences at all observed moments. The two results of model theory that are directly applicable in support of such challenges are *Robinson Consistency Theorem* and *Łoś Theorem.* Robinson Consistency Theorem simply states that the union of two theories is satisfiable under a model if and only if their intersections are consistent, in other words: there is only one interpretation of truth valid in both models. As it is possible that two theories are using different languages and the resulting sentences are not comparable, Łoś Theorem generalizes the idea of expanding a universe through the Cartesian product and defines filters that allow the comparison in a common equivalent representation.

Epistemologically, this is a very significant insight, as terms like interoperability and composability don't need to be redefined and interpreted, but using formal approaches allows to unambiguously address when to models are equivalent, if transformations between model representations and implementations are lossless, etc. What was subject of discussions and expert opinion now becomes subject of mathematical rigor and proof.

## A NEW LOOK AT V&V

The following graphic was published in Sargent (2007) to explain the relations of real world and simulation world with Verification and Validation (V&V).

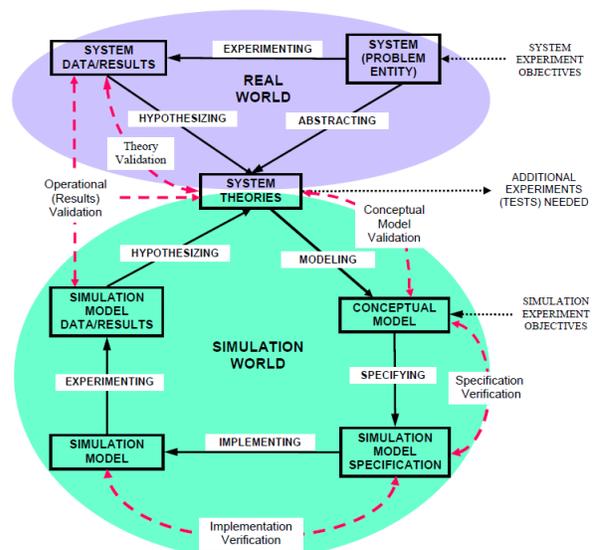

**Figure 1. V&V Relationships (Sargent, 2007)**

Sargent already made clear that we do not use the real system for validation but the system theories that describe our current understanding of this system. The system theories of Isaac Newton's classic physics, Albert Einstein's relativity, Werner Heisenberg's quantum physics, and Brian Greene's string theory are all very different, although all try to describe how the observable world works. When we validate, we validate against such a theory, not the real thing.

Utilizing the insights from the earlier section on model theory, we do not have to rely on subject matter experts and best practices, but we can apply the Robinson Consistency Theorem and Łoś Theorem to proof validity, as long as the system theory is provided in form of a formal language.

In addition, user requirements can now be dealt with in a similar manner: as mathematically rigorously captured data! In systems engineering, e.g., validation actually normally addresses the evaluation if a system meets all user requirements, which means requirements need to be captured accordingly. The system modeling language (SysML) was derived from the universal modeling language (UML) for better support of systems engineering, but added – among some other changed – two important diagrams, namely the *requirement diagrams* that represent a visual modeling of requirements for the system, which are pivotal for systems engineering, and the *parametric diagrams* that show the relations between parameters for system components on all levels. The diagrams allow for tracing of all other classes to the requirements from which they are derived (by relating them to the requirement diagram) as well as introducing metrics to measure the fulfillment of the requirements.

This idea was utilized for the development of the Modeling and Simulation System Development Framework (MS-SDF) described in Tolk et al. (2013) and shown in figure 2.

System theories recognized by the user as well as requirements, assumptions, and constraints are collected in mathematically unambiguous form to make up the *reference model* for the simulation. Tolk et al. (2013) define "*a reference model as an explicit model of a real or imaginary referent, its attributes, capabilities and relations, as well as governing assumptions and constraints under all relevant perceptions and interpretations. The reference model captures what is known and assumed about a problem situation of interest.*" As this model captures all inputs, it is complete regarding defining elements, but likely not consistent, as contradicting system theories are possible and not aligned requirements from different stake holders are highly likely.

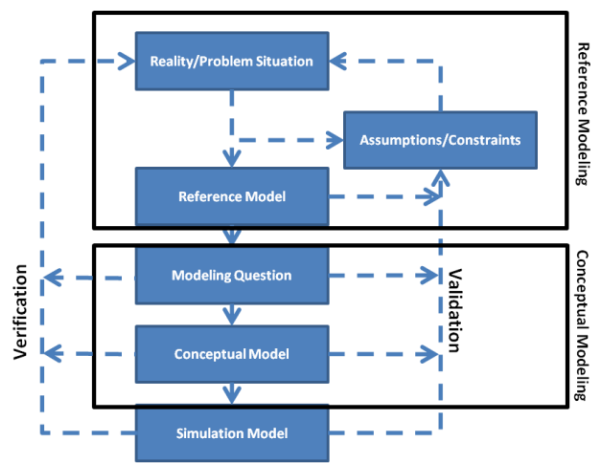

**Figure 2. The M&S System Development Framework (Tolk et al. 2013)**

To build a simulation system, however, we need a consistent conceptual model. Robinson (2008) defines a conceptual model as *"a non-software specific description of the computer simulation model (that will be, is or has been developed), describing the objectives, inputs, outputs, content, assumptions and simplifications of the model."* All this information is captured in the reference model, so that a conceptual model can be extracted by filtering. The filter is defined by the consistency requirement (no contradictions can pass the filter) and the relevance requirement (only concepts identified in the modeling question can pass the filter). In case of contradictions, several conceptual models can be derived that illuminate the different aspects of the challenge as understood in the reference model. This approach was envisioned as multi-simulation in Yilmaz et al. (2007).

A good example for the proposed way is the use of weather models to predict hurricane paths: instead of building one federation of all models, the models are used in parallel and their projections are displayed in a common context, namely as hurricane tracks with wind strength on a map. The common elements shared by all weather models are wind strength and coordinates.

The framework allows identifying common concepts that can be used to display results side-by-side without having to force contradicting theories into one single model. Why this approach, the developing of federations, is epistemologically dangerous will be discussed in our last section.

## DEVELOPING FEDERATIONS

In particular in the military application domain, the use of federations is very common. The idea is intriguing: instead of developing new M&S solutions, existing part solutions are federated to expose their partial functionally so that in the sum the overall need functionality is provided to the user by the federation. If, e.g., a closed air support operation is needed, an air force model can simulate the air crafts while an army model can simulate the land-based operations.

The problem with this approach is that we are implicitly assuming that all participating simulations support one world view. In other words, we are assuming that we start from the common ground of a common and accepted description of reality in form of an object model that can serve as the *Übermodell* from which all simulation representations can be derived by pruning and aggregating. But that is not necessarily true. As Tolk et al. (2011) explain it: *"As we are connecting simulated things we need transparency of what we are simulating, as the real world referent use in other interoperability domains has been replaced in the modeling phase by its representing conceptualization in the M&S interoperability domain."*

This is a general problem. Winsberg (2010) uses Nano sciences as an example where scientists are interested in how cracks evolved and move through material. To address this problem, three different levels of resolution are necessary. In order to understand how cracks begin, sets of atoms governed by quantum mechanics are modeled in small regions. These regions are embedded into medium scale regions that are governed by molecular dynamics. Finally, most of the material is neither cracking nor close to a developing crack and can be modeled using continuum mechanics based on linear-elastic theory. The challenge is that these three theories cannot be mapped to each other as they are inconsistent. No theory exists that allows creating a common theory. However, we can create a federation based on coupling heuristics, which Winsberg calls handshakes. In the given example, the common expression of energy was utilized to exchange information between the regions and allow for a common model that executes perfectly fine. But what is the foundation? This federation is based on three theories that are valid in themselves, but the federation is only based on a heuristic supported by no theory whatsoever, so can it be valid?

## CONCLUSIONS

Let's summarize the findings before coming back to our question posted in the introduction. From the history of science we understand that a theory is the best consistent explanation of observables by providing causality that explains correlations. This causality can be captured by functions and relations. We can capture these in a model that can be implemented as a simulation. However, when the simulation is computer-based, the limitations of computability are given. Model theory helps to better understand when models are consistent with the implemented theory as well as with additional requirements, which is in alignment with ideas of validation and verification, although these processes need to be supported by more rigorous methods. The current practice of building federations ignores or is unaware of these findings and needs to be revisited.

However, does this mean that we cannot gain knowledge from something that is essentially wrong? Weirich (2012) makes a strong case that models are still very useful, as long as we understand the underlying assumptions and constraints. What is needed is a better education of M&S engineers to understand the principles captured in this paper. A simulationists must not only be a good computer engineer, he also needs to understand the underlying conceptualizations and mathematics that are the philosophical foundations on which M&S as a discipline is build. Our current M&S curricula do not all support this well enough, and the need is not yet fully recognized in industry. With this paper, we hope to reach some decision makers to address these issues better in the future.

The second question posted in the introduction, whether simulation stands equally beside theory and experimentation, needs also further discussion (Padilla et al., 2013). In this paper, the authors observe that simulation is either used for theory building, theory testing, or a mixed mode thereof. When simulations are used to generate data from which new insights are derived, this process is the creating of a new theory. If data is compared with empirical evidence, the simulation is tested using the methods of theory testing. If empirical data is used to test a simulation and afterwards the simulation is used to generate new data to be evaluated, we have a mixed form. In all cases, the simulation either stands in place of the theory, or in the place of the experimentation. Again, there is a vast amount of methods applicable that has been developed of centuries of theory and simulation development. Simulationists must know what they are doing and which methods are applicable.

M&S as a discipline is still young and in development. That is why it is even more important to address these questions and embed it into the context of other scientific disciplines.

**ACKNOWLEDGEMENTS**

The authors like to thank their colleague from the Epistemology of Simulation (EPOS) group for the discussions and ideas during the recent workshops.